\documentclass[12pt]{article}
\usepackage{graphicx}
\begin{document} 

\centerline{\bf
Ising, Schelling and Self-Organising Segregation}

\bigskip

\noindent
D. Stauffer$^1$ and S. Solomon

\bigskip
\noindent
Racah Institute of Physics, Hebrew University, IL-91904 Jerusalem, Israel

\bigskip
\noindent
$^1$ Visiting from
Institute for Theoretical Physics, Cologne University, D-50923 K\"oln, Euroland

\bigskip
\centerline{e-mail: stauffer@thp.uni-koeln.de, sorin@cc.huji.ac.il}
\bigskip

Abstract: The similarities between phase separation in physics and 
residential segregation by preference in the Schelling model of 1971
are reviewed. Also, new computer simulations of asymmetric interactions
different from the usual Ising model are presented, showing spontaneous
magnetisation (= self-organising segregation) and in one case a 
sharp phase transition.

\section{Introduction}

More than two millennia ago, the Greek philosopher Empedokles (according to 
J. Mimkes) observed than humans are like liquids: Some mix easily like 
wine and water, and some do not, like oil and water. Indeed, many binary fluid 
mixtures have the property that for temperatures $T$ below some critical 
temperature $T_c$, they spontaneously separate into one phase rich in one of 
the two components and another phase rich in the other component. For $T > T_c$,
on the other hand, both components mix whatever the mixing ratio of the two
components is. Chemicals like isobutyric acid and water, or cyclohexane and 
aniline, are examples with $T_c$ near room temperature, though they smell
badly or are poisonous, respectively. For humans, segregation along racial, 
ethnic, or religious lines, is well known in many places of the world. 

Schelling \cite{schelling} transformed the Empedokles idea into a quantitative
model and studied it.  People 
inhabit a square lattice, where every site has four neighbours to the North,
West, South and East. Everyone belongs to one of two groups A and B and prefers
to be have neighbours of the same group more than to be surrounded by 
neighbours of the other group. Thus with some probability depending on the 
numbers $n_A$ and $n_B$ of neighbours of the two groups, each person moves 
into a neighbouring empty site. After some time with suitable parameters,
large domains are formed which are either populated mostly by group A or
mostly by group B. 
 
Physicists use the Ising model of 1925 to look at similar effects. Again each
site of a large lattice can be A or B or empty; A and B are often called 
``spin up'' and ``spin down'' in the physics literature referring to 
quantum-mechanical magnetic moments. The probability to move depends 
exponentially on the ratio $(n_A-n_B)/T$ calculated from the neighbour states.
A B ``prefers'' to be surrounded by other B, and an A by other A. The lower 
this temperature or tolerance $T$ is the higher is the probability for A to 
move to A-rich neighbourhoods, and for B to move to B-rich neighbourhoods.
Therefore at low $T$ an initially random distribution of equally many A and B 
sites will separate into large regions (``domains'') rich in A, others rich in
B, plus empty regions. In magnetism these domains are called after Weiss 
since a century and correspond to the ghettos formed in the Schelling model.

This effect can be seen easier without any empty sites. Then either a site
A exchanges places with a site B, or a site A is replaced by a site B and
vice versa, where in the above probabilities now $n_A$ and $n_B$ are the
number of A and B sites in the two involved neighbourhoods. Or, even simpler,
a site A changes into a site B or vice versa, involving only one neighbourhood. 
The latter case can be interpreted as an A person moving into another city, 
and another person of type B moving into the emptied residence. The
physics literature denotes the exchange mechanism as Kawasaki kinetics, 
the switching mechanism as Glauber (or Metropolis, or Heat Bath) kinetics.
Again, at low enough $T$ large A domains are formed, coexisting with 
large B domains. In the simpler switching algorithm, finally one of these
domains wins over the other, and the whole square lattice is occupied by 
mostly one type, A or B.

The above $T$ can instead of temperature be interpreted socially
as tolerance: For high $T$ no such segregation takes place and both groups mix 
completely whatever the overall composition is. Instead of ``tolerance'' we may
interpret $T$ also as ``trouble'': External effects, approximated as random 
disturbances, may prevent people to live in the preferred residences, due to
war, high prices, low incomes, pecularities of the location, .... Some of 
these effects were simulated by Fossett \cite{fossett}. 
Without these empty sites, we may also interpret A as one type of liquid and 
B as the other type, and then have a model for the above-mentioned binary 
liquids which may or may not mix with each other via the Kawasaki exchange
of places. Alternatively, we may interpret A as a high-density liquid and B as 
a low-density vapour and then have a model for liquid-vapour phase transitions:
Only below some very cold temperature can air be liquefied. The first 
approximate theory for these liquid-vapour equilibria is the van der Waals 
equation of 1872. 

Thus Schelling could have based his work on a long history of physics research,
or a film of computer simulation published in Japan around 1968. But in 1971 
Schelling did now yet know this physics history \cite{aydin} and his model
was therefore more complicated than needed and was at that time to our
knowledge not yet simulated in the Ising model literature. Schelling did not
consider $T > 0$ and at $T = 0$ his model has problems (see below) with 
creating the predicted segregation. Even today, 
sociologists \cite{clark,fossett,edmonds,zhang} do not cite the physics 
literature on Ising models. Similarly, physics journals until a few years ago
ignored the 1971 Schelling publication \cite{levy}, though recently physicists
extended via Ising simulations the Schelling model to cases with $T$ increasing 
with time \cite{ortmanns} and involving more than two groups of people 
\cite{schulze}. However, applications of the Ising model to social 
questions are quite old \cite{galam}. 

In the following section we point out an artifact in the old Scheling model
and a simple remedy for it, coming from the rule how to deal with people
surrounded by equal numbers of liked and disliked neighbours. 
We explain in the next section in greater detail the standard Ising simulation
methods using the language of human segregation. Then we present two new
models. One takes into account that human interactions, in contrast to 
particles in physics, can be asymmetric: If a man loves a woman it may 
happen that she does not love him, while in Newtonian physics actio = 
--reactio: An apple falls down because Earth attracts the apple and the apple
attracts Earth. The other model introduces holes (empty residences) similar to 
the original Schelling work, with symmetric interactions. Also, we check for 
sharp transitions and smooth interfaces in a Schelling-type model.

\section{Artifact in Schelling model}

In Schelling's 1971 model, each site of a square lattice is occupied by 
a person from group A, or a person from group B, or it is empty. People 
like to have others of the same group among their eight (nearest and 
next-nearest) neighbours and require that ``no fewer than half of one's  
neighbors be of the same'' group (counting only occupied sites as 
neighbouring people). Thus, if a person has as many A as B neighbours,
then in the Schelling model that person does not yet move to another 
site. Imagine now the following configuration with 12 people from group B
surrounded by A on all sides:
\bigskip
\newpage

A A A A A A A A

A A A A A A A A

A A A B B A A A

A A B B B B A A

A A B B B B A A

A A A B B A A A

A A A A A A A A

A A A A A A A A

\bigskip

In this case not a single B has a majority of A neighbours, and all A have 
a majority of A neighbours. Thus none would ever move, and the above
configuration is stable. (Similar artifacts are known from Ising models at zero 
temperature \cite{redner}.) One can hardly regard the above configuration as 
segregation when 8 out of 12 B people have a balanced neighbourhood
of four A and four B neighbours each. And this small cluster
does not grow into a large B ghetto. Also larger configurations with this 
property can be invented. In fact, at a vacancy concentration of 30 \% and
starting from a random distribution our simulations gave only small domains, 
with no major changes after about 10 iterations. 

To prevent this artifact one should in the case of equally many A and B 
neighbours allow with 50 percent probability the person to move to another
place; and we will implement such a probabilistic rule later.

\section{Ising model}

Fossett \cite{fossett} reviews the explanations of segregation by preference
of the individuals or by discrimination from the outside. In Schelling's model 
\cite{schelling}, preference alone could produce segregation, but in reality
also discrimination can play a role. For example, Nazi Germany established
Jewish ghettos by force in many conquered cities. A simple Ising model without
interactions between people can incorporate discrimination with a field $h$.
We assume that a site which is updated in a computer algorithm is occupied
with probability $p_A$ proportional to exp($h$) by a person from group A, and 
with probability $p_B \propto \exp(-h)$ by a B person. Properly normalized we 
have
$$ p_A = e^h/(e^h + e^{-h}) , \quad p_B = e^{-h}/(e^h + e^{-h}) \eqno (1)$$
leading to 
$$ -M = (e^h-e^{-h})/(e^h + e^{-h}) = {\rm tanh}(h)   \eqno (2) $$
for the relative difference $M = (N_B-N_A)/N$ of all A and B people in large
lattices with $N$ sites. There is no need for any computer simulations in this
simple limit without interactions between people. In reality, one may have
a discrimination with positive $h$ in one part of the lattice and negative 
$h$ in the rest of the lattice, leading to segregation by discrimination.

Now we generalize the field to include besides this discrimination $h$ 
also the interactions of site $i$ with its four nearest neighbours, of which 
$n_A$ are of type $A$ and $n_b = 4-n_A$ are of type B:
$$h_i =  (n_A-n_B)/T' + h  \eqno(3)$$
where $T'$ is the tolerance towards neighbours from the other group; now also
the probabilities 
$$ p_A(i) = e^h_i/(e^h_i + e^{-h_i}) , \quad p_B(i) = e^{-h_i}/(e^h_i + e^{-h_i}) 
\eqno (4)$$
depend on the site $i$. This defines the standard Ising model on the square 
lattice; of course many variations have been simulated since around 1960,
and theoretical arguments showed $T_c = 2/\ln(1 + \sqrt2) \simeq 2.2$. Thus for 
all $T'$ below $T_c$ at $h = 0$ the population separates into large B-rich and 
A-rich domains with composition $(1 \pm M)/2$, 
whose size increases towards infinity with time, while for $T'
 > T_c$ no such ``infinitely'' large domains are formed.  Thus we now define
$T = T'/T_c$ such that for $T < 1$ we have segregation and for $T > 1$ we 
have mixing, at zero field.  Schelling starts with random configurations but
then uses more deterministic rules, analogous to $T = 0$.
However, only for $T < 1$ this spatial separation
leads to domains growing to infinity for infinite times on infinite lattices.

For positive $h$, the equilibrium population always has A as majority and B as 
minority. If we start with a A majority but make $h$ small but negative, then 
the system may stay for a long time with an A majority until it suddenly 
``turns'' \cite{fossett} into a stronger B majority: Nucleation in metastable 
states, like the creation of clouds if the relative humidity exceeds 100 
percent (in a pure atmosphere).

(Physicists call the above method the heat bath algorithm; alternatives are
the Glauber and the Metropolis algorithms. The choice of algorithms
affects how fast the system reaches equilibrium and how one specific 
configuration looks like, but the average equilibrium properties are not 
affected. That remains mostly true also if in Kawasaki kinetics these updates 
of single sites are replaced by exchanging the people on two different sites.
In contrast, if the lattice is diluted by adding empty sites as in 
\cite{schelling}, then the transition $T$ may be different from 1.)

Of course, this Ising model is a gross simplification of reality, but these
simplifications emphasise the reasons for spontaneous segregation. As 
stated on page 210 of Fossett \cite{fossett}: ``{\it Any} choice to seek 
greater than proportionate contact with co-ethnics necessarily diminishes the
possibility for contact with out-groups and increases spatial separation
between groups; the particular motivation behind the choice (i.e.,
attraction vs. aversion) may be a matter of perspective and in any case is 
largely beside the point.'' 
 
\begin{figure}[hbt]
\begin{center}
\includegraphics[angle=-90,scale=0.5]{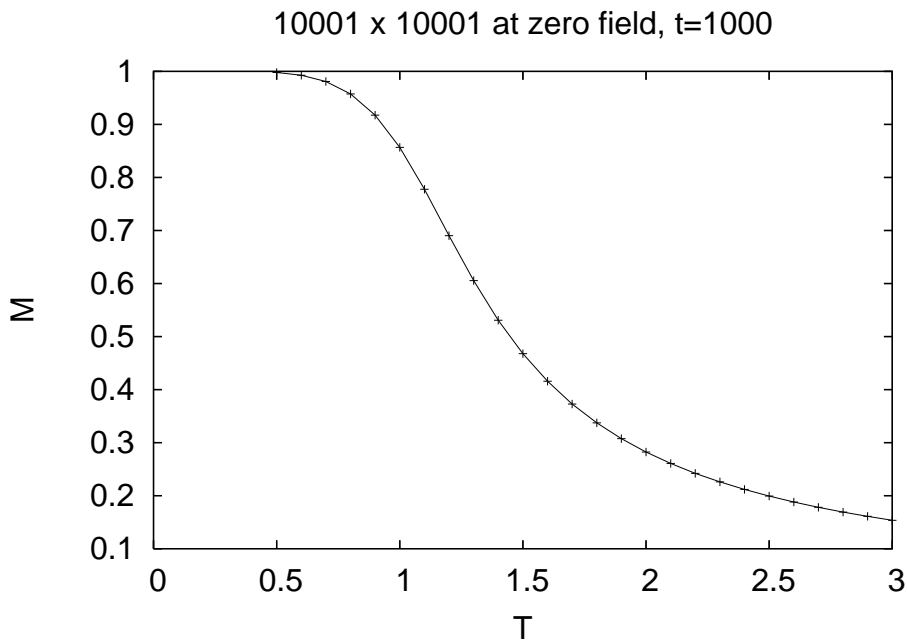}
\end{center}
\caption{Composition of the population  versus $T$ at $h=0$, averaged over
1000 sweeps through a lattice of hundred million people.
}
\end{figure}

\begin{figure}[hbt]
\begin{center}
\includegraphics[angle=-90,scale=0.5]{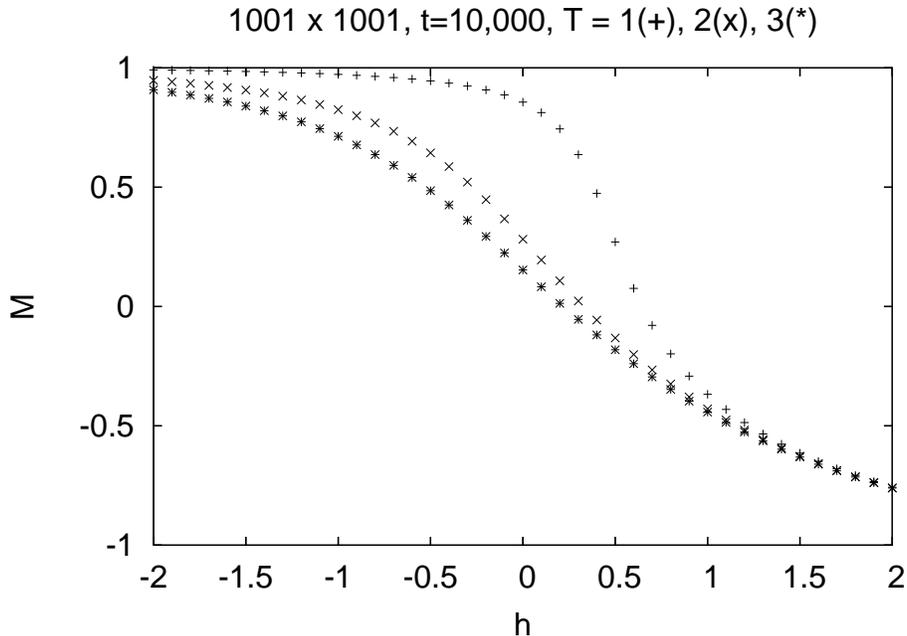}
\end{center}
\caption{Composition of the population versus $h$ at fixed $T = 1,2,3$, 
averaged over 10,000 sweeps through a lattice of one million people. This
simulation took 4 1/2 hours.}
\end{figure}

\section{Modifications}

\subsection{Asymmetric simulations}

In the above model, the rules are completely symmetric with respect to A and B.
Fossett \cite{fossett} reviews the greater willingness of the minority B in 
American racial relations to mix with the majority A, compared with the
willingness of A to accept B neighbours. This we now try to simulate by moving
away from physics and by assuming that A is more influenced by B than B is
influenced by A. Thus if in the above rule, 3 or 4 of the neighbours are 
A, then $p_A(i) = p_B(i) = 1/2$. Mathematically, eq.(3) is replaced by 
$$h_i = \min(0,n_A-n_B)/T + h \eqno (5)$$
in our modification. The neutral case of probabilities 1/2 then occurs if 
A is replaced by B, or B is replaced by A, in a predominantly A neighborhood.

Now the previous sharp transition at $T = 1, \, h=0$ vanishes: Fig.1 shows 
smooth curves of $M$ versus $T$ for $h=0$, and Fig.2 shows smooth curves of 
$M$ versus $h$ at three fixed $T$. Maybe this smooth behaviour is judged more
realistic by sociology. No segregation into large domains happens, and in 
contrast to the symmetric Ising model of the preceding section, the results are 
the same whether we start with everybody A or everybody B. 

\begin{figure}[hbt]
\begin{center}
\includegraphics[angle=-90,scale=0.5]{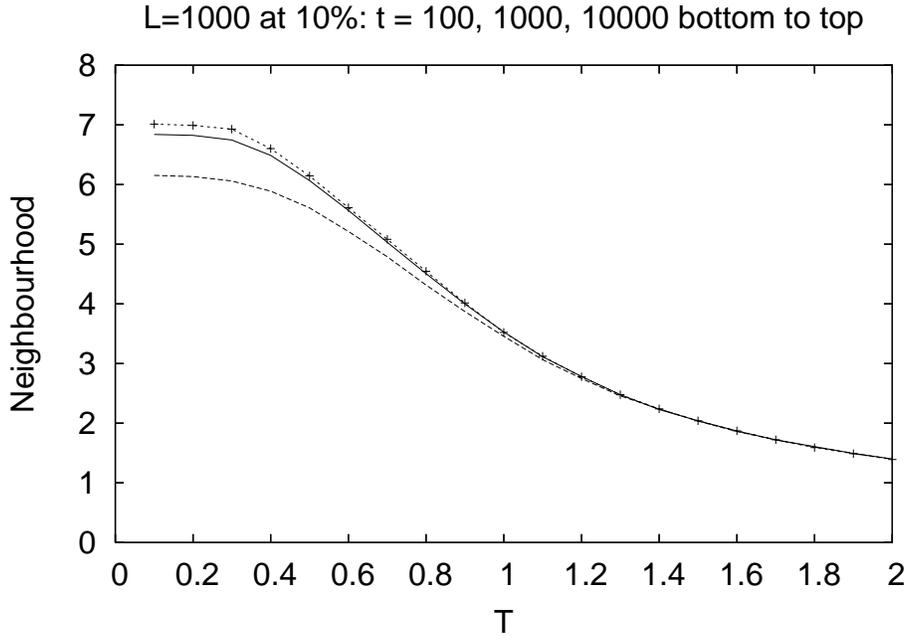}
\end{center}
\caption{$T$ dependence of the average number of same minus different 
neighbours, for three times $t$ showing that about 1000 iterations are 
enough.
}
\end{figure}

\begin{figure}[hbt]
\begin{center}
\includegraphics[angle=-90,scale=0.5]{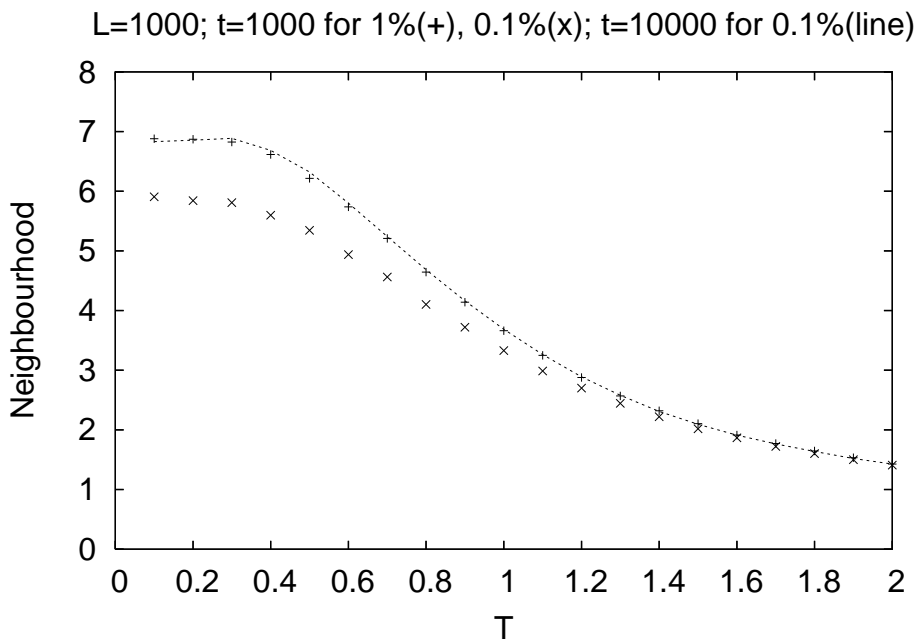}
\end{center}
\caption{As Fig.3 but for vacancy concentrations of 0.1 and 1 \%.
}
\end{figure}

\subsection{Empty spaces}

Schelling had to introduce holes (= empty residences) on his lattices since
he did now allow a B person to become A or vice versa (via moving to another
city) and moved only one person at a time (not letting two people exchange
residences). Now we check if holes destroy the sharp transition between
self-organised segregation and no such segregation. In physics this is called
``dilution'', and if the holes are fixed in space one has ``quenched''
dilution. In this case the fraction of randomly placed holes must stay below
0.407 to give segregation into ``infinitely'' large domains; for larger hole
concentration the lattice separates into fixed finite neighbourhoods of 
people, separated by holes, such that infinite domains are impossible
(``percolation'' \cite{aharony}). For housing in cities, it is more 
realistic to assume that holes are not fixed: An empty residence is occupied
by a new tenant who leaves elsewhere the old residence empty; physicists
call this ``annealed dilution''.

Thus besides A and B sites we have holes (type C) of concentration $x$,
while A and B each have a concentration $(1-x)/2$. People can move into an 
empty site or exchange residences (``Kawasaki kinetics'') with people of
the other group, i.e. A exchanges sites with B.

We also replaced the $n_A-n_B$ in eq.(3) by the changes in the number 
of ``wrong'' neighbours. Thus we calculate the number $\Delta$ of A-B 
neighbour pairs before and after the attempted move, and make this 
move with a probability proportional to exp($-\Delta/T'$); no overall
discrimination $h$ was applied. Thus this symmetric model assumes that A does 
not like to have B neighbours, and B equally does not like A neighbours, while
both do not care whether a neighbouring residence is empty or occupied by
people of the same group.

Now the total number of A, B and C sites is constant, and a quantity like
the above $M$ no longer is useful. Inspection of small lattices shows that
again for low $T$ large domains are formed, while for large $T$ they are not 
formed. To get a more precise border value for $T$, we let A change into B
and B change into A. Then for $T \le 1.2$ we found that one of the two
groups (randomly selected) is completely replaced by the other, while 
for $T \ge 1.3$ they both coexist.

\subsection{Schelling at positive $T$}

Now we simulate a model closer to Schelling's original version, but at
$T > 0$, while Schelling dealt with the deterministic motion at $T = 0$. 
Thus the neighbourhood now includes eight intead of four sites, i.e. besides
the four nearest-neighbours we also include the four next-nearest (diagonal)
neighbours. Let $n_s(i)$ and $n_d(i)$ be at any moment the numbers of same and
different neighbours, respectively, for site $i$, without counting holes, and 
let sign be the function sign(k) =1 for $k > 0$, $=0$ for $k=0$ and $=-1$ for 
$k< 0$. A person at site $i$ has an ``effort'' 
$$ E_i = {\rm sign}[n_d(i)-n_s(i)]    \quad . \eqno(6)$$
Analogously, $E_j$ is based on the numbers of neighbours of the same and the
different type if the person would actually move into residence $j$. 
In Schelling's $T=0$ limit, nobody would move away from $i$ if $E_i < 0$
and nobody would move into an empty site $j$ with $E_j >0$; instead, people
with $E_i > 0$ move into the nearest vacancy $j$ with $E_j \le 0$.

In reality, one cannot always get what one wants and may have to move
into a ``bad'' neighbourhood. Thus at positive ``temperature'' $T$ we assume
that the move from $i$ to $j$ is made with probability 
$$ p(i \to j) = e^{-\Delta/T}/(1 + e^{-\Delta/T}) \eqno(7b)$$
where 
$$\Delta = E_j - E_i \eqno (7b)$$
is the effort the person at site $i$ needs in order to move to the vacancy at site $j$.
For $\Delta > 0$, higher $T$ correspond to higher probabilities to move
against the own wish, while for the Schelling limit $T \to 0$ nobody moves
against the own wish. For negative $\Delta$ one ``gains'' effort and is likely
to make that move, with a probability the higher the lower $T$ is. For 
$T = \infty$ or $\Delta = 0$ the probability to move is 1/2. Each person
trying to move selects randomly a vacancy from an extended neighbourhood 
up to a distance 10 in both directions; after ten unsuccessful attempts to
find any vacancy the person gives up and stays at the old residence during 
this iteration. (We no longer distinguish in this subsection between $T$ and 
$T'$. Note that $E_i$ is not an energy in the usual physics sense, and thus
this model is not of the Ising type.) 

\begin{figure}[hbt]
\begin{center}
\includegraphics[angle=-90,scale=0.32]{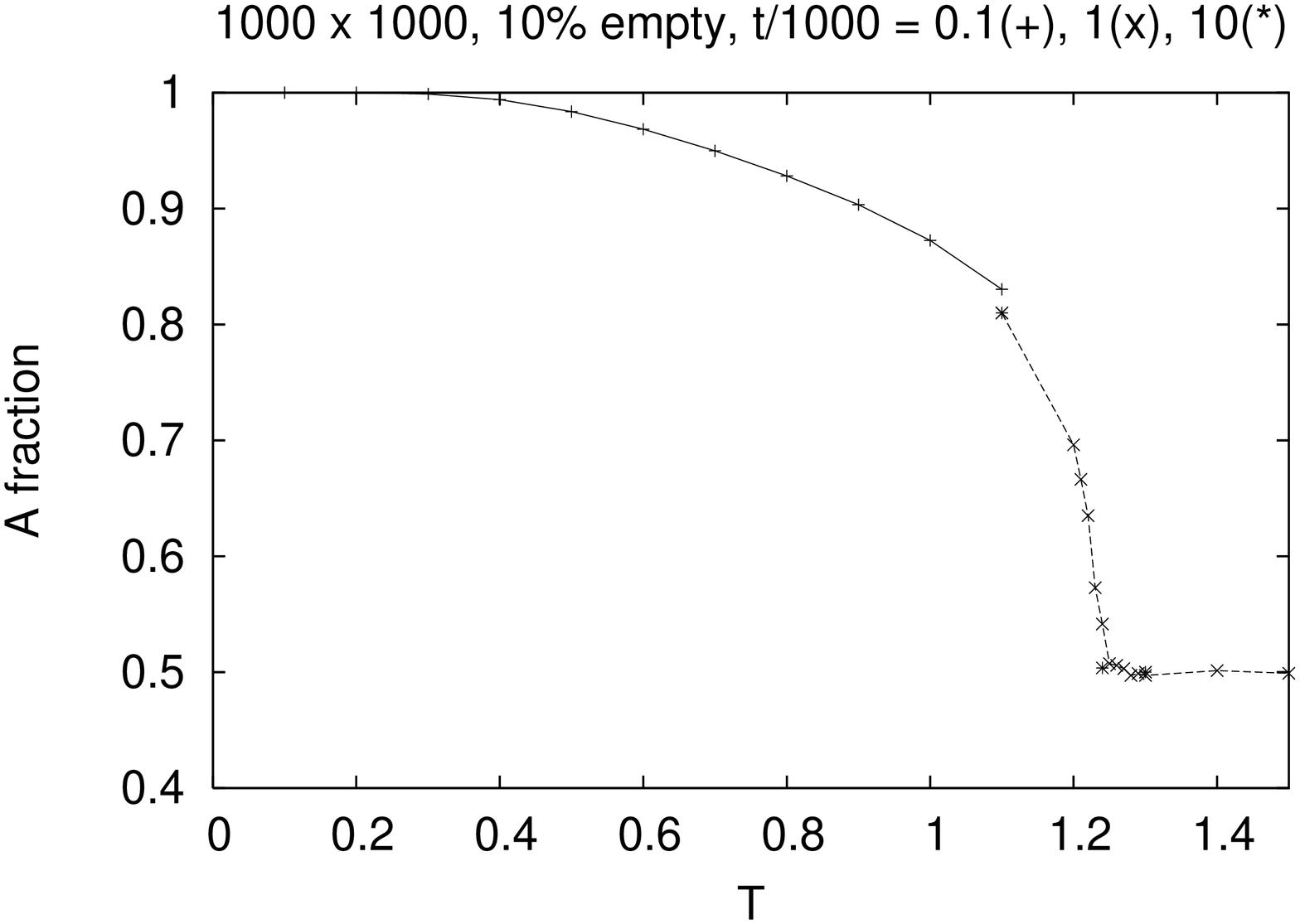}
\includegraphics[angle=-90,scale=0.32]{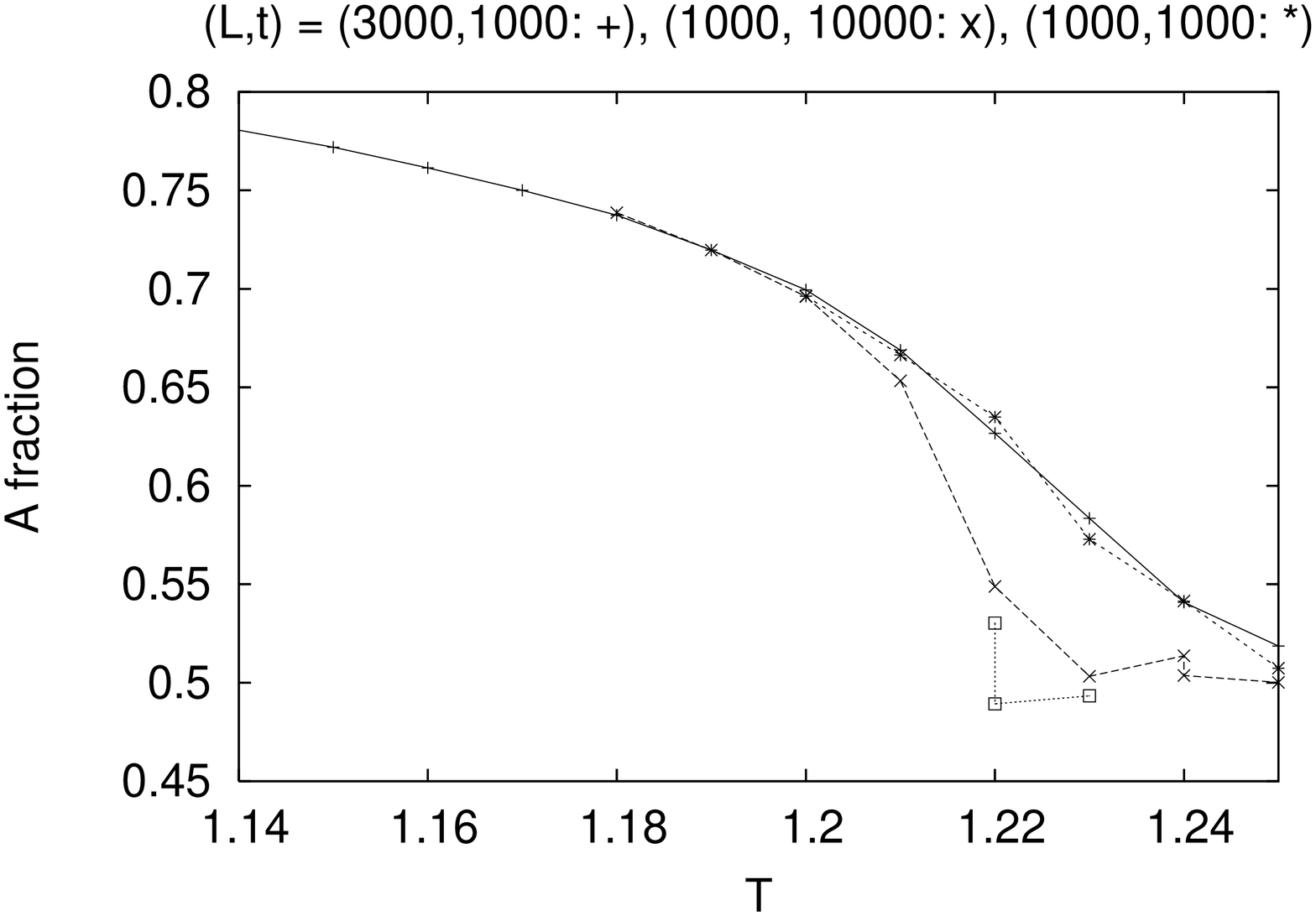}
\end{center}
\caption{Spontaneous A-aggregation, i.e. the self-organising degree of 
segregation $N_A/(N_A+N_B) = (1-M)/2$ versus $T$ in $1000 \times 1000$ lattices after
100 to 10,000 iterations (top). Bottom: additional data up to
$t = 10^5$ (squares) close to $T_c$. 10 percent are vacancies.
}
\end{figure}

\begin{figure}[hbt]
\begin{center}
\includegraphics[angle=-90,scale=0.5]{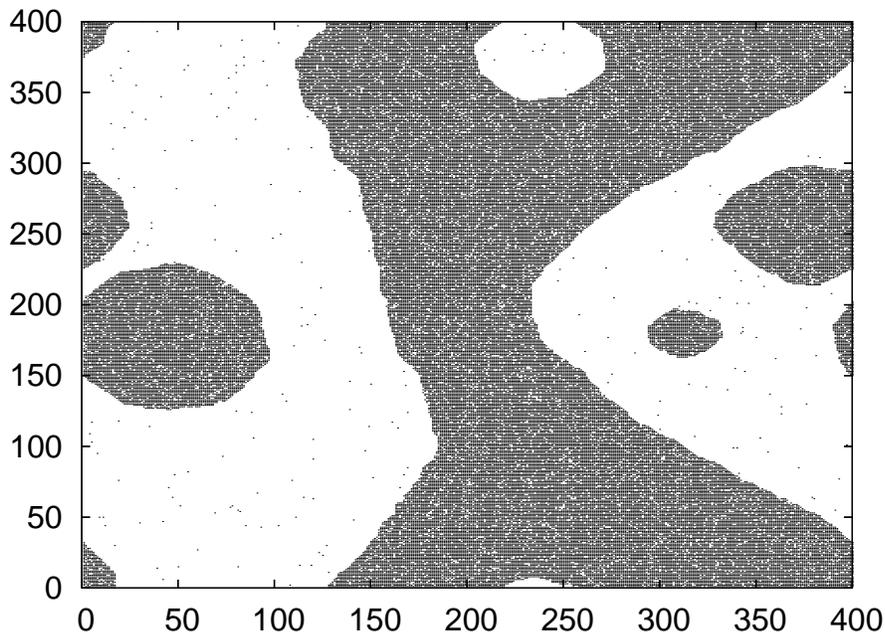}
\end{center}
\caption{Distribution of the A population at $T = 0.1$ after  
100,000 iterations, showing segregation. 10 percent are vacancies.
}
\end{figure}

\begin{figure}[hbt]
\begin{center}
\includegraphics[angle=-90,scale=0.5]{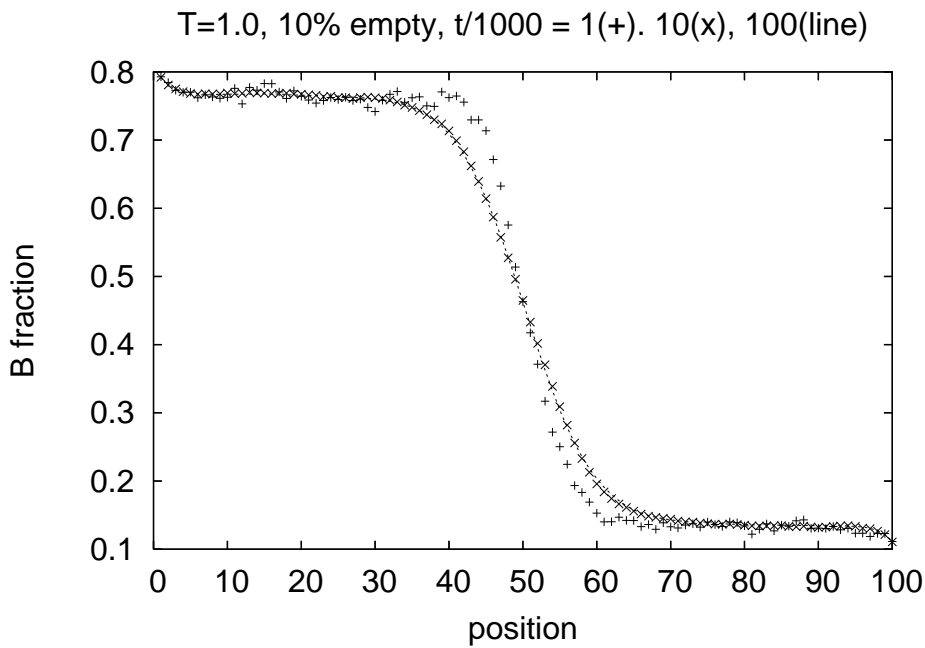}
\end{center}
\caption{Profile of the B fraction as a function of position in a 
$1000 \times 100$ lattice, with the interface between A and B domain
parallel to the longer side of the rectangle. Averaged over the second
half of the simulation.
}
\end{figure}

Figure 3 shows the average ``neighbourhood'' $n_s-n_d$, not counting vacancies, 
for $1000 \times 1000$ lattices for $ t = 100, \, 1000$ and 10,000 iterations
(regular sweeps through the lattice) at a vacancy concentration of 10 \%. 
Already lattices of size $100 \times 100$ agree with Fig.3 apart from minor
fluctuations. Fig.4 shows that for low vacancy concentrations one needs longer
times: At 1 \% and $t = 1000$ the results agree with those at 0.1 \% and 
$t = 10,000$. Although for $T \to 0$ our model does not agree {\textit exactly}
with \cite{schelling} (see Introduction) these figures show clearly the 
Schelling effect at low $T$: A becomes surrounded mainly wth A neighbours and 
B with B neighbours, without any outside discrimination. For large $T$, 
however, this bias becomes much smaller.

Fig.5 shows the overall fraction of group A (ignoring vancancies) in the 
interior of large A-rich domains.  Fig.6 shows partly the time dependence of 
segregation, very similar to standard
Ising model simulations. For low $T$ we see how very small clusters of A sites
increase in size, without yet reaching the size of our $400 \times 400$ 
lattice. In contrast, for high $T$ these clusters do not grow (not shown).
We estimate that near $T = 1.22$ the phase transition occurs between segregating
and not segregating conditions, at a vacancy concentration of 10 percent. 

Starting in the upper half of the system with one group and in the lower half 
with the other group, Fig.7 shows for $T < Tc$  how the interface
between these to initial domains first widens but then remains limited.

\section{Discussion}

The similarities between the Schelling and Ising models have been exploited to
introduce into the Schelling model the equivalent of the temperature $T$.
This turns out to be a crucial ingredient since it ensures that in the presence 
of additional random factors the segregation effect can disappear totally
in a quite abrupt way. Thus cities or neighbourhoods that are currently 
strongly polarized may be transformed into an uniformly mixed area by tiny 
changes in the external conditions: school integration, financial rewards,
citizen campaigns, sport centers, common activities, etc. One-dimensional
models, like some of Schelling's work, are problematic since at positive
$T$ the Ising and many other models do not have a phase transition, while 
they have one in two and more dimensions. 

Besides reviewing the Ising model for non-physicists, we introduced a few 
modifications to it. Together with those of \cite{ortmanns,schulze} they
are only some of the many possible modifications one could simulate. 
Some confirm the result of Schelling, that even without any 
outside discrimination, the personal preferences can lead to 
self-organised segregation into large domains of either mainly A or mainly B 
people. Other modifications or high $T$ (temperature, tolerance, trouble)
prevent this segregation. Thus humans, like milk and honey, are complicated 
but some of their behaviour can be simulated.

The Schelling model is a nice example how research could have progressed better
by more interdisciplinary cooperation between social and natural sciences,
and we hope that our paper helps in this direction.

We thank Maxi San Miguel for sending us \cite{aydin}, and A. Kirman for
discussion.

\end{document}